\documentclass[11pt]{llncs}

\usepackage{amsfonts}
\usepackage{epsfig}

\newcommand{\id}{\;\mbox{$\rm{I} \hspace{-1.0mm}  {\bf I}$}\,}
\newcommand{\ket}[1]{\left\vert#1\right\rangle}
\newcommand{\bra}[1]{\left\langle#1\right\vert}
\newcommand{\minivalmed}[1]{\langle#1\rangle}

\begin{document}

\title{Bypassing state initialisation in perfect state transfer protocols on spin-chains}

\author{C. Di Franco\inst{1} \and M. Paternostro\inst{2} \and M. S. Kim\inst{3}}

\institute{Department of Physics, University College Cork, Cork, Republic of Ireland \and School of Mathematics and Physics, Queen's University, Belfast BT7 1NN, United Kingdom \and Institute for Mathematical Sciences, Imperial College London, SW7 2PG, United Kingdom \\ and \\ QOLS, The Blackett Laboratory, Imperial College London, Prince Consort Road, SW7 2BW, United Kingdom}

\maketitle

\begin{abstract}
Although a complete picture of the full evolution of complex quantum systems would certainly be the most desirable goal, for particular Quantum Information Processing schemes such an analysis is not necessary. When quantum correlations between only specific elements of a many-body system are required for the performance of a protocol, a more distinguished and specialised investigation is helpful. Here, we provide a striking example with the achievement of perfect state transfer in a spin chain without state initialisation, whose realisation has been shown to be possible in virtue of the correlations set between the first and last spin of the transmission-chain.
\end{abstract}

\section{Introduction}
Quantum Information Theory (QIT) is having a remarkable impact from a fundamental viewpoint by providing alternative perspectives to physical problems using new conceptual instruments. The study of quantum correlations shared by many distinctive objects is helping us in understanding their behaviour at critical points~\cite{criticalpoints} and quantifying the resources required in order to efficiently simulate such situations~\cite{simulation}. The simulation of complex quantum systems is usually a prohibitive task for even the most powerful classical machine due to the exponential growth of its Hilbert space with respect to the number of elements. In this context, several advances have recently been made in the study of the ground state of particular many-particle systems~\cite{groundstate}. While all the proposed methods have found use in simulating static properties of ground states, their application to the investigation of time evolution is, in general, problematic. However, although the analysis of the complete behaviour of quantum many-particle systems will be a fundamental task in QIT, for the study of some particular Quantum Information Processing (QIP) schemes this is not necessary. When quantum correlations between only specific elements of a many-body system are required for the performance of a protocol, a more distinguished and specialised approach is helpful. The problems related to simulating the dynamics of many-particle systems can be solved, in this context, by considering not their whole evolution, but only the behaviour of a few characteristic features. Here we provide a striking example in the achievement of perfect Quantum State Transfer (QST) in a spin chain without state initialisation, whose realisation has been shown to be possible in virtue of the correlations set between the first and last spin of the transmission-chain~\cite{roastedchicken}.

This result is also important on a more pragmatic ground. Recently, it has been shown that the control over multipartite registers for the purposes of QIP can be sensibly reduced in a way so as to avoid the generally demanding fast and accurate inter-qubit switching and gating. In this case, the price to pay for the performance of efficient operations is the pre-engineering of appropriate patterns of couplings~\cite{preengineering}. The preparation of a fiducial state for the initialisation of a QIP device can be however experimentally demanding. This is mainly due to the difficulty of preparing pure states of multipartite systems, which is one of DiVincenzo's criteria~\cite{divincenzo}, a set of requirements that any QIP system should meet. Remarkably, our proposal is able to bypass the initialisation of the spin-medium in a known pure state. The scheme requires only end-chain single-qubit operations and a single application of a global unitary evolution and is thus fully within a scenario where the control over the core part of the spin medium is relaxed in favour of controllability of the first and last element of the chain. The relaxation of the conditions necessary for manipulating information is a necessary step in order to shorten the time for the achievement of realistic QIP. This allows us to loose the requirements for information protection from environmental effects. Instead of utilising demanding always-on schemes for the shielding of the information content of a system, this could be done only during the running-time of the protocol.

\section{Perfect state transfer without state initialisation for the $X\!X$ model}
Spin chains have recently emerged as remarkable candidates for the realisation of faithful short-distance transmission of quantum information~\cite{unmodulated}. Here, the system under investigation is a nearest-neighbour $X\!X$ coupling involving $N$ spin-$1/2$ particles. Its Hamiltonian reads
\begin{equation}
 \hat{{\cal H}}=\sum^{N-1}_{i=1}J_{i}(\hat{X}_{i}\hat{X}_{i+1}+\hat{Y}_i\hat{Y}_{i+1}),
\end{equation}
where $J_i$ is the interaction strength between spin $i$ and $i+1$ and $\hat{X},\,\hat{Y}$ and $\hat{Z}$ denote the $x,\,y$ and $z$ Pauli matrix, respectively. Let us start considering
\begin{equation}
J_{i}=J\sqrt{i(N-i)}
\end{equation}
with $J$ being a characteristic energy scale that depends on the specific physical implementation of the model (we choose units such that $\hbar=1$ throughout the paper). This model has been extensively analysed~\cite{cambridge}: $1\rightarrow{N}$ perfect QST is achieved, through this coupling, when the initial state of all the spins but the first one is $\ket{0}$. 
 
\begin{figure}[t]
\centerline{\psfig{figure=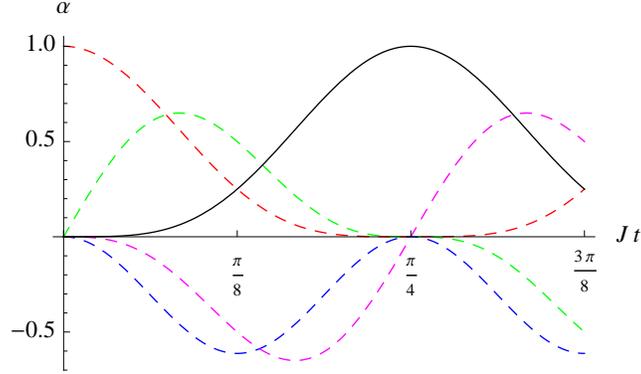,height=5cm}}
\caption{Coefficients $\alpha_1$ (red dashed line), $\alpha_2$ (green dashed line), $\alpha_3$ (blue dashed line), $\alpha_4$ (purple dashed line) and $\alpha_5$ (black line) against dimensionless time $Jt$, for $N=5$ and $J_{i}=J\sqrt{i(N-i)}$.} 
\label{fig2}
\end{figure}

In our investigation, however, we drop the condition on the state of the central qubits, and we just assume control over the external ones.  For the understanding of what follows, it is useful to analyse the time-evolution, in the Heisenberg picture, of the two-site operators $\hat{\id}_i\hat{Z}_{N-i+1}$, $\hat{X}_i\hat{X}_{N-i+1}$, and $\hat{X}_i\hat{Y}_{N-i+1}$. We define $\hat{{\cal O}}(t)$ as the time-evolved form of a given operator $\hat{O}$. By solving a set of Heisenberg equations, we have that, at time $t^*=\pi/4J$ and for any $N$,
\begin{equation}
 \hat{\id}_i(t^*)\hat{\cal Z}_{N-i+1}(t^*)=\hat Z_i\hat{\id}_{N-i+1}.
\label{eq:evolutioncambridgez}
\end{equation}
On the other hand, for an even $N$ we find
\begin{eqnarray}
\nonumber&\hat{\cal X}_i(t^*)\hat{\cal X}_{N-i+1}(t^*)=\hat X_i\hat X_{N-i+1},\\
&\hat{\cal X}_i(t^*)\hat{\cal Y}_{N-i+1}(t^*)=\hat Y_i\hat X_{N-i+1},
\label{eq:evolutioncambridge}
\end{eqnarray}
while for an odd $N$
\begin{eqnarray}
\nonumber&\hat{\cal X}_i(t^*)\hat{\cal X}_{N-i+1}(t^*)=\hat Y_i\hat Y_{N-i+1},\\
&\hat{\cal X}_i(t^*)\hat{\cal Y}_{N-i+1}(t^*)=-\hat X_i\hat Y_{N-i+1}.
\label{eq:evolutioncambridge2}
\end{eqnarray}
These results can also be easily obtained from the analysis presented in Ref.~\cite{ijqi}, where it is shown that the evolution of single-qubit operators can be evaluated by means of a method based on oriented graphs. For instance, in the case $N=5$, $\hat{\cal X}_1(t)$ can be decomposed as
\begin{eqnarray}
\nonumber&\hat{\cal X}_1(t)=\alpha_1(t)\hat{X}_1+\alpha_2(t)\hat{Z}_1\hat{Y}_2+\alpha_3(t)\hat{Z}_1\hat{Z}_2\hat{X}_3+\\
&\alpha_4(t)\hat{Z}_1\hat{Z}_2\hat{Z}_3\hat{Y}_4+\alpha_5(t)\hat{Z}_1\hat{Z}_2\hat{Z}_3\hat{Z}_4\hat{X}_5.
\end{eqnarray}
Similarly,
\begin{eqnarray}
\nonumber&\hat{\cal X}_5(t)=\beta_1(t)\hat{X}_5+\beta_2(t)\hat{Z}_5\hat{Y}_4+\beta_3(t)\hat{Z}_5\hat{Z}_4\hat{X}_3+\\
&\beta_4(t)\hat{Z}_5\hat{Z}_4\hat{Z}_3\hat{Y}_2+\beta_5(t)\hat{Z}_5\hat{Z}_4\hat{Z}_3\hat{Z}_2\hat{X}_1.
\end{eqnarray}
The time-evolution of the two-site operator $\hat{X}_1\hat{X}_5$ can be therefore obtained by considering the sum of all the possible products of elements of these two sets of operator. For $J_{i}=J\sqrt{i(N-i)}$, the time-dependent coefficients $\alpha_i(t)$ have the behaviour shown in Fig.~\ref{fig2}.
By symmetry, $\beta_i(t)=\alpha_i(t)$ for all values of $i$ and $t$. It is easy to notice, in Fig.~\ref{fig2}, that at the time $t^*=\pi/4J$, $\alpha_5(t^*)=\beta_5(t^*)=1$, while all the other coefficients are equal to $0$. For that particular time, therefore, the evolved operator $\hat{\cal X}_1(t^*)\hat{\cal X}_5(t^*)$ is just the product of $\hat{Z}_1\hat{Z}_2\hat{Z}_3\hat{Z}_4\hat{X}_5$ times $\hat{Z}_5\hat{Z}_4\hat{Z}_3\hat{Z}_2\hat{X}_1$. We have $\hat{\cal X}_1(t^*)\hat{\cal X}_5(t^*) = \hat Y_i\hat Y_{N-i+1}$. In the same way, all the other evolved operators in Eqs.~(\ref{eq:evolutioncambridgez}) and (\ref{eq:evolutioncambridge2}) can be obtained.

\begin{figure}[t]
\centerline{\psfig{figure=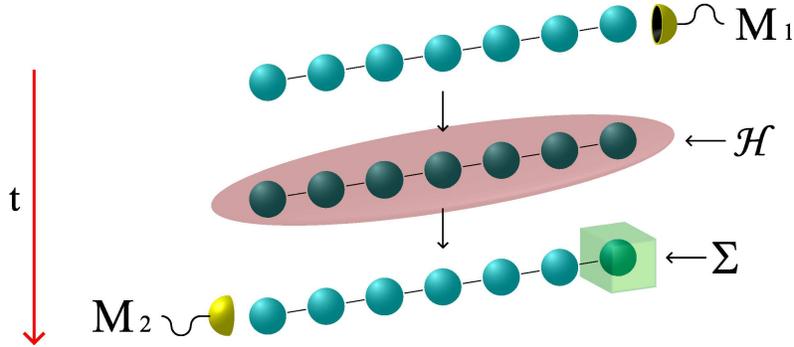,height=5cm}}
\caption{Sketch of the scheme for perfect QST. $M_1$ and $M_2$ are measurements performed over a fixed basis and $\Sigma$ is a conditional operation. $\hat{\cal H}$ is the Hamiltonian of the system.} 
\label{fig3}
\end{figure}

Each of the two-site operators in Eqs.~(\ref{eq:evolutioncambridgez})-(\ref{eq:evolutioncambridge2}) evolves into operators acting on the same qubits, without any dependence on other operators of the chain. This paves the way to the core of our protocol, which we now describe qualitatively. Qubit $1$ is initialised in the input state $\rho^{in}$ (either a pure or mixed state) we want to transfer and qubit $N$ is projected onto
\begin{equation}
\ket{\pm_N}=\frac{1}{\sqrt{2}}(\ket{0}\pm e^{iN\frac{\pi}{2}}\ket{1}).
\end{equation}
In what follows, we say that outcome $+1$ ($-1$) is found if a projection onto $\ket{+_N}$ ($\ket{-_N}$) is performed. Then the interaction encompassed by $\hat{\cal H}$ is switched on for a time $t^*=\pi/4J$, after which we end up with an entangled state of the chain. The amount of entanglement shared by the elements of the chain depends critically on their initial state. Regardless of the amount of entanglement being set, an $\hat{X}$-measurement over the first spin projects the $N$-th one onto a state that is locally-equivalent to $\rho^{in}$. More specifically, if the product of the measurement outcomes at $1$ (after the evolution) and $N$ (before the evolution) is $+1$ ($-1$), the last spin will be in $(\hat{T}^N)^\dag\rho^{in}(\hat{T}^N)$ [$\hat{Z}(\hat{T}^N)^\dag\rho^{in}(\hat{T}^N)\hat{Z}$], where $\hat{T}=\ket{0}\bra{0}+e^{i\frac{\pi}{2}}\ket{1}\bra{1}$ (therefore, $\hat{T}^2=\hat{Z}$)~\cite{comment}. In any case, apart from a simple single-spin transformation, perfect QST is achieved. A sketch of the scheme is presented in Fig.~\ref{fig3}.

The crucial point here is that, regardless of the amount of entanglement established between the spin-medium and the extremal elements of the chain ({\it i.e.} spins $1$ and $N$), upon $\hat{X}$-measurement of $1$, the last spin is disconnected from the rest of the system, {\it whose initial state is inessential to the performance of the protocol} and could well be, for instance, a thermal state of the chain in equilibrium at finite temperature. In fact, the key requirements for our scheme are the arrangement of the proper time-evolution (to be accomplished within the coherence times of the system) and the performance of clean projective measurements on spin $1$ and, preventively, on $N$.

\section{General conditions}
In general, the protocol can be adapted to any Hamiltonian for which we can find a triplet of single-spin operators $\hat{\cal B},\hat{C},\hat{D}$ such that, for symmetric spin pairs, we have
\begin{equation}
\hat{\cal B}^{j_{O}}_i(t^*)\hat{C}_{N-i+1}\hat{\cal O}_{N-i+1}(t^*)=\hat{O}_{i}\hat{D}^{k_O}_{N-i+1}.
\end{equation}
 Here, $\hat{\cal B}_{i}$ ($\hat{D}_{N-i+1}$) provides the eigenbasis for the measurement over spin $i$ ($N-i+1$) of the chain after (before) the evolution, $\hat{C}_{N-i+1}$ is a decoding operation, $\hat{O}_i=\hat{\cal O}_i(0)=\hat{X},\hat{Y},\hat{Z}$ and $j_O,k_O=0,1$, depending on the coupling model. We point out that, when these conditions are not fulfilled, our protocol can still be rather successful. In these cases, through an information flux approach, we can still estimate the average transfer fidelity~\cite{informationflux}. For instance, we can consider the case in which we are able to engineer the strength of the coupling rates of just the extremal qubits ($J_1$ and $J_{N-1}$). Therefore, we take $J_i = J$ (for ${i}=2,...,N-2$), $J_1 = J_{N-1} = \eta{J}$. The behaviour of this system against the dimensionless interaction time $J t$ and the inhomogeneity parameter $\eta$ has already been studied in Ref.~\cite{informationflux}. For simplicity, here we consider the time-dependent coefficients $\alpha_i(t)$ in the case $N=5$, for the value of $\eta$ which maximises QST fidelity ($\eta\sim0.815$). Also this system is centro-symmetric, therefore we have $\beta_i(t)=\alpha_i(t)$ for all values of $i$ and $t$. In this case, however, there is no time for which $\alpha_5(t)=\beta_5(t)=1$, while all the other coefficients are equal to $0$. Nevertheless, for a dimensionless time $J t'\sim1.9$, the value of  $\alpha_5$ and $\beta_5$ are close to $1$, while all the other ones are close to $0$. Our estimate gives an average transfer fidelity via our protocol of $F\sim\alpha_5^2(t')>99.9\%$.

\section{Remarks}
We have shown the existence of a simple control-limited scheme for the achievement of perfect QST in a system of interacting spins without the necessity of demanding state initialisation. Our protocol requires just {\it one-shot} unitary evolution and end-chain local operations. Its efficiency arises from the establishment of {\it correlations} between the first and last spin of the transmission-chain. With the exception of limiting cases where the transfer is automatically achieved (as for the transfer of eigenstates of $\hat{X}_1$ when model $\hat{\cal H}$ is used), these are set regardless of the state of the spin medium, their amount being a case-dependent issue. The end-chain measurements, which are key to our scheme, ``adjust'' such correlations in a way so as to achieve perfect QST.

Due to the dependence of this protocol only on the correlations established between the first and last spin, the analysis of this scheme just requires the investigation of the time evolution of two-site operators. The exponential growth of the Hilbert space of the total state of the system does not affect our analysis, that can thus be done by means of only ``slowly-growing'' computational effort. In fact, the number of elements in the decomposition of the relevant two-site operators grows as $N^2$. Moreover, in this way, we were able to obtain our results removing any dependance on the state of all the central qubits. 

We would like to conclude this contribution by remarking that our protocol for state transfer without initialisation is already encountering the attention of the community interested in quantum spin-chain dynamics. In fact, a recent proposal by Markiewicz and Wiesniak~\cite{MW} has addressed a scheme for perfect state transfer without initialisation where the necessity for ``remote coordination'' between sending and receiving agents is bypassed.

\section{Acknowledgments}
We acknowledge support from the UK EPSRC. C.D.F. is supported by the Irish Research Council for Science, Engineering and Technology. M.P. thanks the UK EPSRC (EP/G004579/1) for financial support.


\begin{thebibliography}{000}
\bibitem{comment} The final state of spin $N$ can be easily obtained from Eqs.~(\ref{eq:evolutioncambridgez})-(\ref{eq:evolutioncambridge2}). For instance, if $N$ is even, $\minivalmed{\hat{Z}_N(t^*)}=\minivalmed{\hat{Z}_1(0)}$, $\minivalmed{\hat{X}_1\hat{X}_N(t^*)}=\minivalmed{\hat{X}_1\hat{X}_N(0)}$ and $\minivalmed{\hat{X}_1\hat{Y}_N(t^*)}=\minivalmed{\hat{Y}_1\hat{X}_N(0)}$. If qubit $N$ has been projected onto $\ket{\pm_N}$ [for which $\minivalmed{\hat{X}_N(0)}=\pm(-1)^{\frac{N}{2}}$], we have $\minivalmed{\hat{Z}_N(t^*)}=\minivalmed{\hat{Z}_1(0)}$, $\minivalmed{\hat{X}_1\hat{X}_N(t^*)}=\pm(-1)^{\frac{N}{2}}\minivalmed{\hat{X}_1(0)}$ and $\minivalmed{\hat{X}_1\hat{Y}_N(t^*)}=\pm(-1)^{\frac{N}{2}}\minivalmed{\hat{Y}_1(0)}$. The state of spin $N$, after the measurement performed on spin $1$, will satisfy $\minivalmed{\hat{Z}_N(t^*)}=\minivalmed{\hat{Z}_1(0)}$, $\minivalmed{\hat{X}_N(t^*)}=(-1)^{\frac{N}{2}}c\,\minivalmed{\hat{X}_1(0)}$ and $\minivalmed{\hat{Y}_N(t^*)}=(-1)^{\frac{N}{2}}c\,\minivalmed{\hat{Y}_1(0)}$, where $c$ is the product of the measurement outcomes at $1$ (after the evolution) and $N$ (before the evolution). The state $(\hat{Z}^{\frac{N}{2}})\rho^{in}(\hat{Z}^{\frac{N}{2}})$ [$(\hat{Z}^{\frac{N}{2}+1})\rho^{in}(\hat{Z}^{\frac{N}{2}+1})$] satisfies these conditions for $c=1$ ($c=-1$).
\end{thebibliography}
\end{document}